\documentstyle[aps,prb,epsf,psfig,floats,twocolumn]{revtex}
\input epsf.tex

\begin{document}

\twocolumn[\hsize\textwidth\columnwidth\hsize\csname@twocolumnfalse%
           \endcsname

\draft

\title{ Exactly solvable model of\\  
        dissipative vortex tunneling}

\author{Akakii Melikidze}
\address{Physics Department, Princeton University, Princeton, NJ 08544}

\date{\today}

\maketitle

\begin{abstract}
I consider the problem of vortex tunneling in a two-dimensional
superconductor. The vortex dynamics is governed by the Magnus force and
the Ohmic friction force. Under-barrier motion in the vicinity of the  
saddle point of the pinning potential leads to a model with quadratic
Hamiltonian which can be analytically diagonalized. I find the
dependence of the tunneling probability on the normal state quasiparticle
relaxation time $\tau$ with a minimum  at $\omega_0\tau\sim 1$, where
$\omega_0$ is the level spacing of the quasiparticle bound states
inside the vortex core. The results agree qualitatively with the available
experimental data.
\end{abstract}


\vskip 0.3 truein
]

%%%%%%%%%%%%%%%%%%%%%%%%%%%%%%%%%%%%%%%%%%%%%%%%%%%%%%%%%%%%%%%%%%%%%%%%%

\section{Introduction}

The dynamics of vortices in Type-II superconductors has been one of the
most attractive areas of research in recent years~\cite{Bla} because of
its importance from scientific as well as technological point of
view. To science, vortices present an example of extended topologically
stable objects with extremely rich properties. On the technological side,
the motion of vortices is the source of supercurrent dissipation - the
circumstance that limits high magnetic fields application of high-$T_c$
materials. In view of this fact, the exploration of the limits of strong
pinning of vortices has been actively pursued both in theory and 
experiment. It has long been known that the thermally-activated depinning
rate strongly decreases as the temperature is lowered and eventually
saturates at values believed to be set by quantum tunneling. The present
article explores this quantum regime.

The difficulty of the problem is in the fact that there are two major
forces that govern the dynamics of the vortex: Hall (Magnus) force and
friction force. While the first one is conservative and can be treated
through a single vortex description, the second one is not: the energy
dissipates into the environment. Therefore, one needs to consider the
combined vortex-plus-environment system. In the pioneering paper by
Feigel'man {\it et al.},~\cite{Fei} which took into account both forces,
the environmental degrees of freedom were integrated out with the use
of path-integral techniques to produce an effective vortex action 
description. The time non-locality of the resulting action, however,
limited the analysis to qualitative conclusions. Later, the same approach
was independently taken by Morais Smith {\it et al.},~\cite{Mor} but they
also failed to go beyond the scaling analysis of the effective action.

This paper is devoted to the study of an exactly solvable model of
two-dimensional dissipative vortex tunneling. It has a quadratic
Hamiltonian of the vortex coupled to the environment and is solvable by an
analytic diagonalization. The present model is a generalization of
the quadratic model of one-dimensional dissipative particle tunneling
that was solved by Ford {\it et al.}~\cite{ForLewOco} The results of the
solution show the dependence of the tunneling rate on the ratio of the
Magnus and dissipative forces. The tunneling rate has a minimum near the
point where these two forces become equal.

This paper is organized as follows: I first start with a
non-dissipative case in Sec.~2; the dissipative model is introduced in
Sec.~3, where some of its general features are analyzed; in Sec.~4 I
describe the solution of the dissipative tunneling problem and compare the
results with available experimental data; conclusions
follow in Sec.~5. Units with $\hbar=1$ are used throughout the paper.

%%%%%%%%%%%%%%%%%%%%%%%%%%%%%%%%%%%%%%%%%%%%%%%%%%%%%%%%%%%%%%%%%%%%%%%%%%

\section{Vortex tunneling: non-dissipative case}

In order to analyze the motion of a vortex one needs first to identify the
forces that govern its dynamics. Both in classical~\cite{Saf} and in
quantum~\cite{Don} fluids there is an intrinsic Magnus force that acts on
a vortex when it moves relative to a (super)fluid. This force is normal
both to the vortex tangent and to the vortex velocity, relative to the
fluid, and is linear with in the latter. In this respect, it is very
similar to Lorentz force that acts on charged particles moving in magnetic
fields. Since Magnus force does not produce any work, it is possible to
formulate vortex dynamics using Hamiltonian formalism.~\cite{Saf,Don} This
formalism was successfully applied to the problem of
quantum-mechanical vortex nucleation by Volovik.~\cite{Vol} However, it
was not until much later that Haldane and Wu~\cite{HalWu} realized that,
just like Lorentz force is a manifestation of a geometric
(Aharonov-Bohm) phase~\cite{ShaWil} associated with a motion of a
particle, Magnus force should arise as a consequence of a geometric phase
associated with vortex motion. Later, Ao and Thouless~\cite{AoTho}
extended this idea to vortices in superconductors. To clarify the physics
involved, I will sketch the derivation of the geometric phase done by Ao
and Thouless.

Consider the ground state many-body wave function of a two-dimensional
superconductor: $\Psi_0({\bf r}_1,\ldots,{\bf r}_N)$, where ${\bf
r}_j=(x_j,y_j)$ are the coordinates of electrons. A vortex at ${\bf
r}=(x,y)$ can be created by ``spinning up'' the system:
\begin{eqnarray}
\label{}
\Psi_v = \exp\left\{\frac{i}{2}\sum_{j=1}^{N}
             \theta({\bf r}_j-{\bf r})\right\}
         \Psi_{0}^{'}({\bf r}_1,\ldots,{\bf r}_N;{\bf r}).
\end{eqnarray}
Here it is assumed that $\Psi_{0}^{'}$ is ``close'' to $\Psi_0$, {\it
i. e.}, it is obtained from the latter by a continuous deformation;
$\theta({\bf r}_j-{\bf r})=\arctan\left(\frac{x_j-x}{y_j-y}\right)$. Note
that the factor $\frac 12$ in the exponent comes from the fact that the
condensate is made of Cooper pairs. Now one
can adiabatically move ${\bf r}$ around a closed contour keeping the
coordinates of electrons fixed. The phase accumulated at the end will come
from the exponent and will be given by $\pi$ times the number of electrons
encircled by the path of the vortex. This phase can be described by the
inclusion of a geometric phase term in the action:
\begin{eqnarray}
\label{S_T}
S = \int dt\, \alpha \dot{x}(t)y(t).
\end{eqnarray}
Here $\alpha=\pi n_s$ is the Hall coefficient and $n_s$ is the density of
electrons in the condensate. One sees that the length scale, associated
with this geometric phase term is $1/\sqrt{n_s}$ which is one of the
smallest in the problem. One expects then that the dynamics will be
dominated by the Magnus force.

Next I include the vortex potential energy in the action:
\begin{eqnarray}
\label{S}
S = \int dt \left[ \alpha \dot{x}y - V(x,y) \right].
\end{eqnarray}
To clarify the picture I go over to the Hamiltonian formulation. First of
all, it is easy to identify canonically conjugate variables. Indeed, from
the above action one sees that the variable, conjugate to $x$ is $\alpha 
y$. The Hamiltonian is then equal to the potential energy:
\begin{eqnarray}
\label{Ham}
H &=& V(x,y), \\
\left[ x,y \right] &=& \frac{i}{\alpha}.
\end{eqnarray}
For an arbitrary $V(x,y)$ there is a problem of operator
ordering. This problem has been addressed by Girvin and
Jach;~\cite{GirJac} an alternative approach to this problem is the
path-integral approach considered by Jain and Kivelson.~\cite{JaiKiv} In
the model studied below, however, this difficulty does not appear.

A comment should be made about the relevance of vortex mass which is not
included in the model. Using the analogy between a superconducting vortex
and a particle in magnetic field (i. e. similar origins of the
corresponding geometric phases) this omission of the vortex mass amounts
to the effective Lowest Landau Level approximation. Such an approximation
is warranted if the transitions to higher Landau levels can be
neglected. This can be checked by comparing matrix elements $|V_{nm}|$ of
the potential energy $V(x,y)$ between different Landau levels with the
spacing $\omega_c$ between Landau levels. The ratio $|V_{nm}|/\omega_c$
depends on the particular potential under consideration and can be
estimated in our problem of vortex tunneling to be $\sim 2\pi/S$, where
$S$ is the action involved in tunneling. The value of this action can
be extracted from experimental data (to be discussed in Sec.~4) and is
$\approx 50$. Thus, the ratio $|V_{nm}|/\omega_c\sim 0.1$ and the omission
of the vortex mass is indeed warranted. However, it is possible that in
some other situations the mass of a vortex is relevant and should be
included. This can be easily done since the mass term in the Hamiltonian
is quadratic and thus does not lead to any substantial complications.

%~~~~~~~~~~~~~~~~~~~~~~~~~~~~~~~~~~~~~~~~~~~~~~~~~~~~~~~~~~~~~~~~~~~~~~~~
\begin{figure}
\epsfxsize=\columnwidth
\centerline{\epsffile{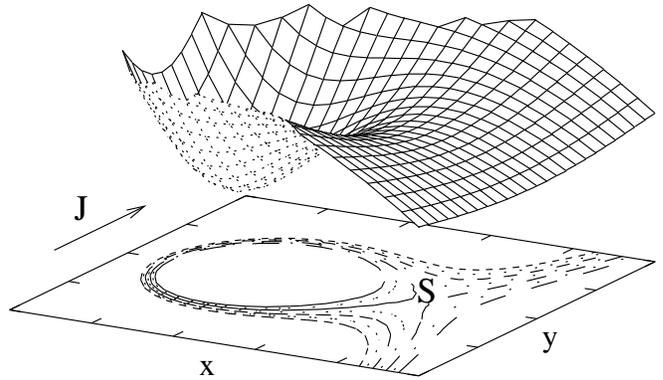}}
\medskip
\caption{Saddle point $S$ in the vortex pinning potential biased by a
          supercurrent $J$ (arbitrary units).}   
\label{Fig1}
\end{figure}
%~~~~~~~~~~~~~~~~~~~~~~~~~~~~~~~~~~~~~~~~~~~~~~~~~~~~~~~~~~~~~~~~~~~~~~~~

Consider now the tunneling problem in which an impurity potential, which
pins the vortex, is tilted by a Magnus force due to an applied depinning
supercurrent (see Fig.~\ref{Fig1}). For a  vortex to tunnel out of the
pinning site it has to overcome a potential barrier. This leads to a
thermally-activated depinning rate at high temperatures. Here I am
considering the problem of quantum tunneling which gives dominant
contribution to the depinning rate at low temperatures. One can argue then
that for sufficiently strong depinning supercurrents the tunneling
exponent is dominated by the under-barrier motion in the vicinity of the
saddle point of the potential (see Fig.~\ref{Fig1}). This leads one to
consider the problem of tunneling in the following model potential:
\begin{eqnarray}
\label{saddle}
V = \frac{\kappa_y}{2} y^2 - \frac{\kappa_x}{2} x^2, 
   \qquad \kappa_x,\kappa_y > 0.
\end{eqnarray}
This becomes the problem of tunneling across an inverted parabolic barrier
$-\kappa_x x^2$ in 1D quantum mechanics if one identifies the momentum
$p=\alpha y$, and the mass $m=\alpha^2/\kappa_y$ of the 1D particle. The
transmission coefficient for this problem can be calculated exactly and is
given by:~\cite{LanLif}
\begin{eqnarray}
\label{D_0}
D_0 (E) &=& \frac{1}{1+\exp\left\{2\pi\frac{E}{\Omega_0}\right\}}, \\
\label{Omega_0}
\Omega_0 &=& \sqrt{\frac{\kappa_x}{m}}
         = \frac{\sqrt{\kappa_x\kappa_y}}{\alpha}.
\end{eqnarray}
Here $-E$ is the energy of the vortex as measured from the value at the
saddle point. $E$ equals the activation energy in the high temperature
thermally-assisted tunneling regime.

In comparison, the previous works on the dissipative vortex tunneling
problem~\cite{Fei,Mor} considered the model with an added cubic term:
$V = \frac{\kappa_y}{2} y^2 - \frac{\kappa_x}{2} x^2 - \frac{\kappa_3}{3}
x^3$ which has a minimum at $x=-\frac{\kappa_x}{\kappa_3}$. I argue that 
the role of the cubic term is mainly to set the position of the localized
state and its energy and that it is not essential for the
tunneling itself. Meanwhile, setting $\kappa_3=0$ produces a quadratic  
Hamiltonian which allows for an exact diagonalization when the dissipation
is added. Finally, let me mention that the model
Eq.~(\ref{saddle}) (without dissipation) has been considered before by    
Fertig and Halperin~\cite{FerHal} in a different context: the tunneling of
electrons in two dimensions in the presence of a strong magnetic field.
It later became the basis for the Chalker-Coddington model of the Integer
Quantum Hall transition.~\cite{ChaCod}

The result Eqs.~(\ref{D_0},\ref{Omega_0}) shows an interesting feature of
the vortex tunneling problem. If $\kappa_y$ is decreased while both
$\kappa_x$ and the height of the barrier $E$ are kept constant the
transmission coefficient $D_0$ decreases implying stronger pinning. In
particular, for $\kappa_y=0$ the vortex no longer tunnels across the
barrier. This result can be understood by observing that for $\kappa_y=0$
the Hamiltonian becomes invariant with respect to the $y$ coordinate of
the vortex. This leads to the conservation of the conjugate momentum
which is $\alpha x$. Thus, the tunneling process in which $x$ changes is
forbidden by a new conservation law. This means that the pinning
potentials which are translationally invariant {\it along} the direction
of the biasing supercurrent are good candidates for strong vortex pinning.

%%%%%%%%%%%%%%%%%%%%%%%%%%%%%%%%%%%%%%%%%%%%%%%%%%%%%%%%%%%%%%%%%%%%%%%%%%

\section{Dissipation}

So far we have considered a vortex in a pure superconductor. In reality,
the presence of impurities that sets the normal state resistance at low
temperatures introduces a finite quasiparticle scattering time
$\tau$. This fact has to be taken into account and the adiabatic arguments
presented above need to be modified. These modifications have been worked
out by Kopnin and Kravstsov~\cite{KopKra} and by Kopnin and
Salomaa~\cite{KopSal} (for a simple new look at the problem see
also~\cite{Sto}). They showed that a finite relaxation time breaks the
adiabaticity in the spectral flow of the quasiparticle vortex core bound
states~\cite{Core_States} and leads to two main effects: the
Hall coefficient $\alpha$ is reduced from its pure value $\pi n_s$, and
there appears a friction force acting on the vortex, ${\bf F}=-\eta {\bf
v}$. The coefficients $\alpha$ and $\eta$ are given by:
\begin{eqnarray} 
\label{variation}
\alpha = \pi n_s \frac{(\omega_0\tau)^2}{1+(\omega_0\tau)^2},\qquad
\eta = \pi n_s \frac{\omega_0\tau}{1+(\omega_0\tau)^2}.
\end{eqnarray}
Here, $\omega_0$ is the level spacing of quasiparticle bound states inside
the
vortex core. Thus, the parameter $\omega_0\tau$ sets relative importance
of the Magnus and friction forces. This can be quantified by defining the
Hall angle: $\tan \Theta_H = \alpha/\eta = \omega_0\tau$. Thus, only in
the ``superclean limit'' $\omega_0\tau\gg 1$ can one neglect the
dissipative force and consider a pure quantum problem.

It should be mentioned that there is an alternative point of
view expressed by Ao {\it et al.}~\cite{AoTho} according to which
the Hall
coefficient is a topological number and thus is not renormalized.
It turns out that the analysis of experimental results of van Dalen
{\it et al.}~\cite{Dal} presented in Sec.~4 suggests that there is indeed
a renormalization of the Hall coefficient $\alpha$. Meanwhile, I take
$\alpha$ and $\eta$ as two phenomenological parameters.

In the following I use the approach of
Ford {\it et al.}~\cite{ForLewOco} to the problem of dissipation in
quantum mechanics. I model the friction force by a linear coupling to a
bath of oscillators (to simplify notation I use $x_a$,
$a=\left\{1,2\right\}$ for $x$ and $y$ coordinates of the vortex):
\begin{eqnarray}
\label{model}
H = V(x_a) + \sum_{aj}\left[ \frac{p_{aj}^2}{2m_j}
                        + \frac{1}{2}m_j\omega_j^2(q_{aj}-x_a)^2
                   \right], \\
\left[x_1,x_2\right] = \frac{i}{\alpha} , \quad
\left[q_{ai},p_{bj}\right] = i\delta_{ab}\delta_{ij}.
\end{eqnarray}
Vortex coordinates $x_1$ and $x_2$ are coupled to two independent
identical sets of oscillators in a simple way: they shift their
equilibrium positions. The difference between this model the one
considered by Ford {\it et al.}~\cite{ForLewOco} is that, due to a
different physical context, here both $x_1$ and $x_2$ need to be coupled
to the oscillator baths in order to account for friction in both
directions.

The Heisenberg equations of motion corresponding to
Eq.~(\ref{model}) are:
\begin{eqnarray}
\label{Heisenberg}
\alpha \epsilon_{ab}\dot{x}_b
   = - \frac{\partial V}{\partial x_a}
       + \sum_j m_j\omega_j^2(q_{aj}-x_a), \\
\label{Heisenberg_2}
\ddot{q}_{aj} +\omega_j^2q_{aj} = \omega_j^2 x_a.
\end{eqnarray}
The solution of Eq.~(\ref{Heisenberg_2}) is:
\begin{eqnarray}
\label{Solution}
q_{aj}(t) = q_{aj}^h(t) + x_a(t) -
\int\limits_{-\infty}^t dt'\,\cos[\omega_j(t-t')]\dot{x}_{a}(t'),
\end{eqnarray}
where $q_{aj}^h(t)$ is the solution of the corresponding homogeneous
equation. Substituting
Eq.~(\ref{Solution}) into Eq.~(\ref{Heisenberg}) one obtains:
\begin{eqnarray}
\label{Effective}
\alpha \epsilon_{ab}\dot{x}_b
   &=& - \frac{\partial V}{\partial x_a}
  -\int\limits_{-\infty}^t dt'\,\mu(t-t') \dot{x}_{a}(t') + N_a(t),\\
\mu(t) &=& \theta(t)\sum_j m_j\omega_j^2\cos(\omega_j t), \\
N_a(t) &=& \sum_j m_j\omega_j^2q_{aj}^h(t).
\end{eqnarray}
Here $\mu(t)$ is the so-called memory kernel, $N_a(t)$ is the noise
force and $\theta(t)$ is the Heaviside step function.

Next I consider the so-called Ohmic case when the friction force acting
on the vortex is strictly linear in the vortex velocity. This is realized
by taking the following distribution of oscillator strengths:
\begin{eqnarray}
\label{Ohmic}
\eta(\omega) = \frac{\pi}{2}\sum_j m_j\omega_j^2
               \left[\delta(\omega-\omega_j)
                     +\delta(\omega+\omega_j)\right]
             = {\rm const}.
\end{eqnarray}
With this choice Eq.~(\ref{Effective}) becomes:
\begin{eqnarray}
\label{Ballance}
\alpha \epsilon_{ab}\dot{x}_b
   &=& - \frac{\partial V}{\partial x_a}
     - \eta \dot{x}_{a} + N_a(t).
\end{eqnarray}
Naturally, this can be called the force balance equation: it
balances Magnus, potential, friction and noise forces. This equation shows
that the model Eq.~(\ref{model}) indeed describes friction force linear in
the velocity. A more general version of this equation is known as the
quantum Langevin equation.~\cite{ForLewOco}

%%%%%%%%%%%%%%%%%%%%%%%%%%%%%%%%%%%%%%%%%%%%%%%%%%%%%%%%%%%%%%%%%%%%%%%%%%

\section{Dissipative vortex tunneling}

Returning to our problem of tunneling in the vortex depinning by an
external supercurrent, one now has to consider Eq.~(\ref{model}) with $V$
given by Eq.~(\ref{saddle}). The crucial observation due to Ford {\it et
al.}~\cite{ForLewOco} (made for a similar problem) is that the
resulting system is a set of coupled oscillators with one of them having a
negative spring constant ($-\kappa_x$). This leads to a spectrum which
contains an isolated mode with purely imaginary frequency $i\Omega^*$.
After the diagonalization of the Hamiltonian one obtains an oscillator
with a (shifted) negative spring constant that is decoupled from the rest
of the oscillators (all with positive spring constants). It is this new
negative spring constant that determines the tunneling amplitude. The
transmission coefficient is then given by Eq.~(\ref{D_0}) with a
substitution $\Omega_0 \to \Omega^*$:
\begin{eqnarray}
\label{D}
D (E) &=& \frac{1}{1+\exp\left\{2\pi\frac{E}{\Omega^*}\right\}}.
\end{eqnarray}
The value of the imaginary eigenmode frequency $i\Omega^*$ can be found by
diagonalizing the Hamiltonian; a much easier way is to observe that this
eigenmode should satisfy the force balance equation,
Eq.~(\ref{Ballance}), without the noise term ({\it i. e.}, the
corresponding equation for the expectation values; it should be stressed
that noise term can be separated due to the linearity of the
equations). This leads to the following equation on $\Omega^*$:
\begin{equation}
\label{Det}
\det \left(
\matrix{
-\kappa_x + \eta \Omega^* & \alpha\Omega^* \cr
-\alpha\Omega^* & \kappa_y + \eta \Omega^*
}
\right) = 0.
\end{equation}

It is easy to understand this equation qualitatively using the following
simple arguments: roughly, all oscillators in the bath (see
Eq.~(\ref{model})) can be divided into two types, those with frequencies
smaller and bigger than $\Omega^*$. Oscillators with small frequencies do
not follow the motion of the vortex and each one effectively increases the
values of $\kappa_y$ and $-\kappa_x$ by $m_j\omega_j^2$. Those with
high frequencies adjust to the vortex motion and thus do not contribute to
the increase of $\kappa$'s. Then a simple estimate gives :
\begin{equation}
\matrix{
\label{Estimate}
\kappa_y &\to& \kappa_y + \sum\limits_{\omega_j\le\Omega^*}m_j\omega_j^2
   & \approx & \kappa_y + C\eta\Omega^*,\cr
-\kappa_x &\to& -\kappa_x + \sum\limits_{\omega_j\le\Omega^*}m_j\omega_j^2  
   & \approx & -\kappa_x + C\eta\Omega^*,
}
\end{equation}
with $C\sim 1$. This indeed agrees qualitatively with the correct
Eq.~(\ref{Det}).

Eqs.~(\ref{D},\ref{Det}) constitute the solution of the problem.
In the rest of this section the implications of this solution for the
existing experimental data of van Dalen {\it et al.}~\cite{Dal} will be
discussed. In these experiments quantum dynamical relaxation rate of
magnetization $Q$ was measured in oxygen depleted thin films of ${\rm
YBa_2Cu_3O_x}$ as a function of the oxygen content $x$. Changing $x$
changes the normal state quasiparticle relaxation time $\tau$
which, in turn, is expected to change in the Hall and friction
coefficients. I use expressions in Eq.~(\ref{variation}) to compare the
results of the solution of the present model with the experiment.

%~~~~~~~~~~~~~~~~~~~~~~~~~~~~~~~~~~~~~~~~~~~~~~~~~~~~~~~~~~~~~~~~~~~~~~~~
\begin{figure}
\epsfxsize=\columnwidth
\centerline{\epsffile{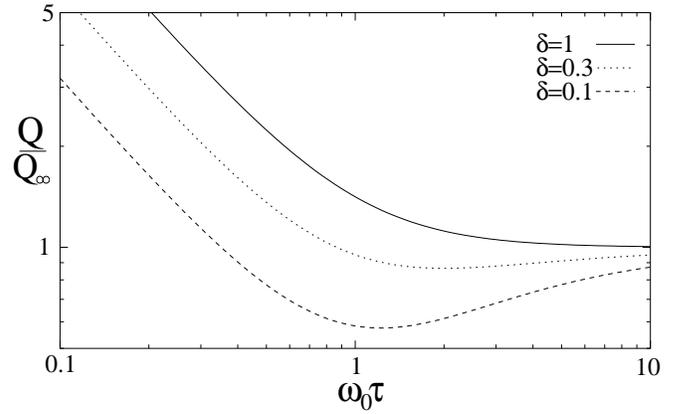}}
\medskip
\caption{Magnetization relaxation rate $Q$ (normalized by its value
$Q_\infty$ at $\omega_0\tau\to\infty$) vs. $\omega_0\tau$ plotted for
various values of $\delta=\kappa_x/\kappa_y\sim\sqrt{1-J/J_c}$.}
\label{Fig2}
\end{figure}
%~~~~~~~~~~~~~~~~~~~~~~~~~~~~~~~~~~~~~~~~~~~~~~~~~~~~~~~~~~~~~~~~~~~~~~~~

The dynamical zero temperature magnetization relaxation rate is $Q=1/S$
where $S$ is the action involved in a tunneling event. It determines the
transmission coefficient of the pinning barrier as $D\sim\exp(-S)$. In
reality $Q$ is small: $Q_\infty\equiv Q(\omega_0\tau\to\infty)\approx {\rm 
0.02}$, therefore Eq.~(\ref{D}) implies: $Q = \Omega^*/2\pi E$. As the
oxygen content, and thus $\omega_0\tau$, is varied $\Omega^*$ (given by
the solution of Eqs.~(\ref{Det},\ref{variation})) changes while $E$ (the
height of the pinning barrier) stays constant. In Fig.~\ref{Fig2} the
calculated ratio $Q/Q_\infty$ is plotted as a function of $\omega_0\tau$
for several values of $\delta = \kappa_x/\kappa_y$. A simple calculation
gives that $\delta\sim\sqrt{1-J/J_c}$, where $J_c$ is the critical biasing
current density at which the minimum of the pinning potential disappears.

From Eq.~(\ref{variation}) one can see that the Hall angle is
$\tan\Theta_H = \alpha/\eta = \omega_0\tau$, therefore in the
``superclean limit'' $\omega_0\tau\to\infty$ the dynamics is dominated by
the Magnus force, while in the opposite ``dirty limit'' $\omega_0\tau\to 0$
it is dominated by the force of friction. Fig.~\ref{Fig2} shows that in
the superclean limit $Q$ saturates while in the dirty limit it varies as
$1/(\omega_0\tau)$. Qualitatively, both of these behaviors were known
before.~\cite{BlaGesVin,Fei} What was not known, however, was the behavior
of $Q$ in
the crossover region $\omega_0\tau\sim 1$. The new qualitative feature of
the exact solution of the present model is a minimum of $Q$ in the
crossover region; it is more pronounced for stronger depinning
currents. This result implies that the strongest pinning occurs
in the ``moderately clean'' regime when $\omega_0\tau$ is of order 1.

In the experiments of van Dalen {\it et al.}~\cite{Dal} only the regime
$\omega_0\tau\le 1$ was realized. There $Q$ shows roughly a
$1/(\omega_0\tau)$ dependence. In the region $\omega_0\tau\sim 1$ the
onset of a crossover in $Q$ is visible, but the minimum
of $Q$ can hardly be observed since no data are available in the
superclean limit $\omega_0\tau\gg 1$. In view of this, experiments on
cleaner films would be desirable to check the existence of the minimum in
$Q$ - the main qualitative predictions of the above analysis.

Finally, it should be emphasized that the $1/(\omega_0\tau)$ behavior of
$Q$ in the dirty limit that was observed in the experiment is a clear
evidence in favor of the correctness of Eq.~(\ref{variation}). Indeed, if
the Hall coefficient $\alpha$ were not renormalized by disorder it would
imply that $Q$ in the dirty limit should always be smaller than its
superclean limit value $Q_\infty$ due to the general tendency of
dissipation to suppress tunneling.~\cite{CalLeg} The experimental results
of van Dalen {\it et al.}~\cite{Dal} show that this is not the case.

%%%%%%%%%%%%%%%%%%%%%%%%%%%%%%%%%%%%%%%%%%%%%%%%%%%%%%%%%%%%%%%%%%%%%%%%%%

\section{Conclusion}

In this article, I have considered a model of dissipative vortex
tunneling. In this model the system-plus-environment Hamiltonian is
quadratic and can be analytically diagonalized. The results were
obtained for the dynamical magnetization relaxation rate $Q$ as a function
of the Hall angle $\tan\Theta_H = \alpha/\eta = \omega_0\tau$, where
$\omega_0$ is the level spacing of the quasiparticle vortex core bound
states. The results show a $1/(\omega_0\tau)$ dependence of $Q$ in the
``dirty'' limit $\omega_0\tau\ll 1$, saturation in the ``superclean''
limit $\omega_0\tau\gg 1$ and a minimum at $\omega_0\tau\sim 1$. These
predictions were compared with the available experimental data of van
Dalen {\it et al.}.~\cite{Dal} Results agree qualitatively in the
``dirty''
regime $\omega_0\tau\ll 1$. On the other hand, the predicted minimum can
not be found since no data are available in the ``superclean'' regime
$\omega_0\tau\gg 1$.

Although I have only considered here the case of a vortex in a
two-dimensional film, the results can be carried over to the
three-dimensional case as well. Following the arguments of
Brandt,~\cite{Bra} one can show that the effect of the extra elastic term in
the energy of a vortex line is mainly to set a length scale along the
vortex - the size of the tunneling nucleus. Meanwhile, the effect of
dissipation on vortex tunneling should be qualitatively the same.

%%%%%%%%%%%%%%%%%%%%%%%%%%%%%%%%%%%%%%%%%%%%%%%%%%%%%%%%%%%%%%%%%%%%%%%%%%

\acknowledgments

I want to thank Duncan Haldane for guidance and criticism and Grigory
Volovik, David Huse and Boris Altshuler for discussions. This work was
supported by NSF MRSEC DMR-9809483.

{\it Note added}. After the initial submission of the article I
became aware of the work by Kim and Shin~\cite{KimShi} in which a similar
problem was treated by a variational method and qualitatively similar
results were obtained. I would like to thank Uwe R. Fischer for referring
me to this work.

%%%%%%%%%%%%%%%%%%%%%%%%%%%%%%%%%%%%%%%%%%%%%%%%%%%%%%%%%%%%%%%%%%%%%%%%%%

\end{document}